# Light rays and waves on geodesic lenses

Lin Xu [1, *], Xiangyang Wang [2, *], Tomáš Tyc [3, *, †], Chong Sheng [2], Shining Zhu [2], Hui Liu [2, ‡] and Huanyang Chen [1, §]

[1] Institute of Electromagnetics and Acoustics and Department of Electronic Science, Xiamen University Xiamen 361005, China

[2] National Laboratory of Solid State Microstructures and School of Physics, Collaborative Innovation Center of Advanced Microstructures, Nanjing University, Nanjing, Jiangsu 210093, China

[3] Department of Theoretical Physics and Astrophysics, Masaryk University, Kotlarska 2, 61137 Brno, Czech Republic

**Starting from well-known absolute instruments for perfect imaging, we introduce a type of rotational-symmetrical compact closed manifolds, namely geodesic lenses. We demonstrate that light rays confined on geodesic lenses are closed trajectories. While for optical waves, the spectrum of geodesic lens is (at least approximately) degenerate and equidistant with numerical methods. Based on this property, we show a periodical evolution of optical waves and quantum waves on geodesic lenses. Moreover, we fabricate two geodesic lenses in sub-micrometer scale, where curved light rays are observed with high accurate precision. Our results may offer a new platform to investigate light propagation on curved surfaces.**

Absolute instruments (AIs) in optics means devices that brings stigmatically an infinite number of light rays from a source to its image, which can perform perfect imaging in the perspective of geometrical optics[1, 2]. Two well-known examples of AIs are a plane mirror and Maxwell's fisheye lens (with gradient refractive index profile, see in Fig. 1(a)). Actually, there are a lot of AIs, such as Eaton lens, Luneburg lens, invisible lens and so on[3]. Recently, one of the authors proposed a general method to design AIs with the help of the Hamilton-Jacobi equation[4], which has flourished the family of AIs. No matter how perfectly stigmatic the geometrical-optics image might be, in the wave-optics regime the resolution is always limited by diffraction. Owing to this limitation, "perfect imaging" in the perspective of geometrical optics and wave optics is quite different[3]. As far as we know, the only AI in both perspectives is Pendry's slab[5]. However, the frequency spectrum of other AIs has been investigated by numerical method[6, 7] and the Wentzel–Kramers–Brillouin (WKB) approximation[8, 9]. It is found that their spectrum is (at least approximately) degenerate and equidistant, which contribute to periodical evolution of waves in AIs [6, 8].

With a conformal coordinate transformation, AIs with cylindrical-symmetry refractive index profiles in two dimensional (2D) space can be connected to curved surfaces with rotational-symmetry embedding in three dimensional (3D) space in the perspective of geometrical optics[10, 11]. Such surfaces are usually called geodesic lenses (GLs) corresponding

---

to the AIs. For the compact GLs, all light rays follow closed trajectories, which correspond to the perfect imaging of AIs[12]. However, for wave optics it has not been clear so far whether the spectrum of GLs shares the properties of spectra of AIs, namely whether it is (at least approximately) degenerate and equidistant or not.

In this paper, we firstly introduce GLs and confirm that the spectrum of compact GLs are (at least approximately) degenerate and equidistant with numerical method. After that, we show a periodical evolution of optical waves on GLs based on this property. Then we make an analogy of quantum waves and demonstrate the periodical evolution on GLs. Finally, we fabricate two GLs in sub-micrometer scale, where closed curved light trajectories are observed with high precision.

**Geodesic lenses constructed from Absolute instruments**

AIs with cylindrical-symmetry refractive index profile denoted by $n(r)$ has been widely studied[3, 11, 12]. We present a Maxwell's fish-eye lens[13], a generalized Maxwell's fish-eye lens[3] with M=5, an extended invisible lens[14] and an inverse invisible lens[12] with contour plots in Fig. 1(a-d), respectively. Note that an invisible lens is a finite device while an extended or inverse one is infinite. The center of each lens is marked with **S**. The position vector is denoted with **r**. Point sources at position **A** have either a perfect image at position **B** (see in Fig. 1(a) and (b)) or have a self-image at position **A** (see in Figs. 1(c) and (d)). By using Hamilton's equation of optics, we can find all the trajectories in AIs as shown in red and green closed curves[15]. There is a conformal coordinate transformation between AIs and GLs[11], written as

$$\rho = n(r)r, \mathrm{d}h = n(r)\mathrm{d}r \qquad (1)$$

where $\rho$ is radial coordinate, and $h$ is the length measured along the meridian from North pole of the GLs. Equation (1) enables to construct a GL from a given index profile of an AI. We have constructed four such GLs that correspond to the above AI index profiles and are shown in Fig. 1(e-h): the sphere, the spindle[12] (we use this name according to its shape), Tannery's pear[16] and truncated Tannery's pear[17], respectively. The list of these GLs along with the corresponding AIs is given in Table 1.

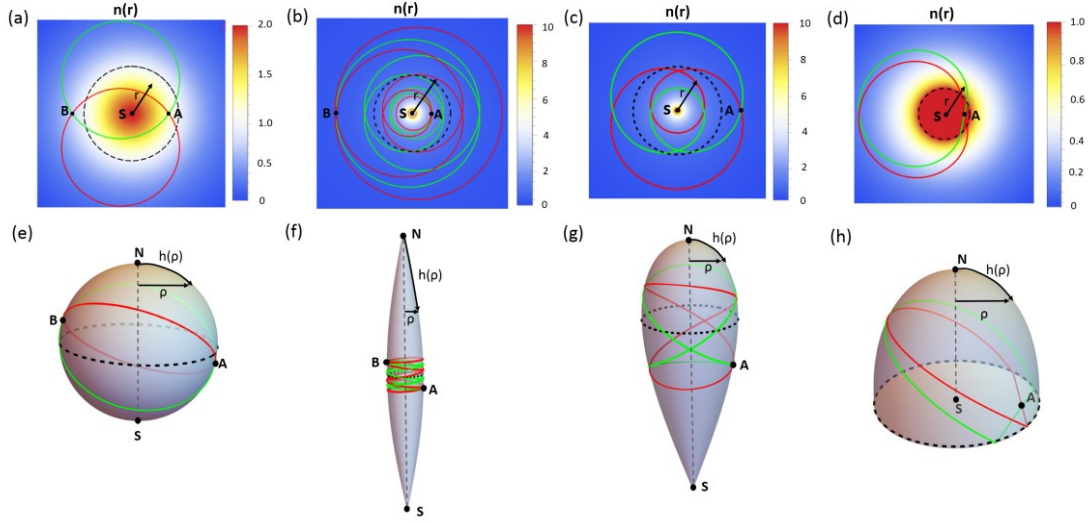

**Figure 1 | AIs (upper row) and corresponding GLs (lower row) with rotational symmetry.** In AIs, the center of each lens is marked with **S**. The position vector is denoted with **r**. The angle parameter $\theta$ is omitted for simplification because of rotational symmetry of AIs and GLs. Contour plots show refractive index profile of $n(r)$. The presented AIs are Maxwell's fish-eye lens (a), generalized Maxwell's fish-eye with M=5 (b), extended invisible lens (c) and inverse invisible lens (d), respectively. The corresponding geodesic lenses are sphere (a), spindle with M=5 (b), Tannery's pear (c) and truncated Tannery's pear (d). On GLs, the axis of rotational symmetry is denoted by a dashed line, which connects North (N) pole and South (S) pole. $\rho$ is the radial coordinate, and $h(\rho)$ is the length measured along the meridian from N pole on the geodesic surface. Light rays starting from point A form closed trajectories shown in different colors. **S** poles are mapped from the centers **S** of AIs, while **N** poles correspond to the infinities of AIs. Dashed black circles of AIs are places with refractive index of unity at radius of $r_0$, which are equivalent to those of GLs.

**Light rays and waves on Geodesic lenses**

An important thing is that the local refractive index profile on GLs is constant (and equal to unity), which is quite different from that of AIs. Light trajectories on GLs can be simply obtained by Eq. (1) from those in the corresponding AIs. Since we know the shape of GLs, we can also find their geodesics by solving the geodesic equations, namely,

$$\frac{d^2 x^\lambda}{d\xi^2} + \Gamma^\lambda_{\mu\nu} \frac{dx^\mu}{d\xi} \frac{dx^\nu}{d\xi} = 0, \quad (2)$$

where $\xi$ is parameter of geodesics, $\Gamma^\lambda_{\mu\nu}$ is the Christoffel symbols of coordinate system $\{x^\lambda; \lambda = 1, 2\}$. Those geodesics are shortest paths in GLs, which correspond to light trajectories. Because of rotational-symmetrical $\rho(h)$ of GLs, we find that the Christoffel symbols are written as,

$$\Gamma^{\lambda}_{\mu\nu} = \left( \begin{pmatrix} 0 & 0 \\ 0 & -\rho(h)\rho'(h) \end{pmatrix}_{\mu\nu} \begin{pmatrix} 0 & \dfrac{\rho'(h)}{\rho(h)} \\ \dfrac{\rho'(h)}{\rho(h)} & 0 \end{pmatrix}_{\mu\nu} \right)_{\lambda}, \quad (3)$$

where we are using the coordinates $\{h, \theta\}$ for $\{x^{\lambda}; \lambda = 1, 2\}$. All the geodesic equations in Eq. (2) are,

$$\begin{cases} h''(\xi) - \rho[h(\xi)]\rho'[h(\xi)]\theta'(\xi)\theta'(\xi) = 0 \\ \theta''(\xi) + 2\dfrac{\rho'[h(\xi)]}{\rho[h(\xi)]} h'(\xi)\theta'(\xi) = 0 \end{cases}. \quad (4)$$

By solving Eq. (4) with $\rho(h)$ and given initial conditions, we can find their geodesics shown in Figs. 1(e-h) with red and green closed curves.

Since the transformation of Eq. (1) is conformal, it preserves the angle of two light trajectories in AIs to that of their images on GLs[12], which might find application in conformal transformation optics.

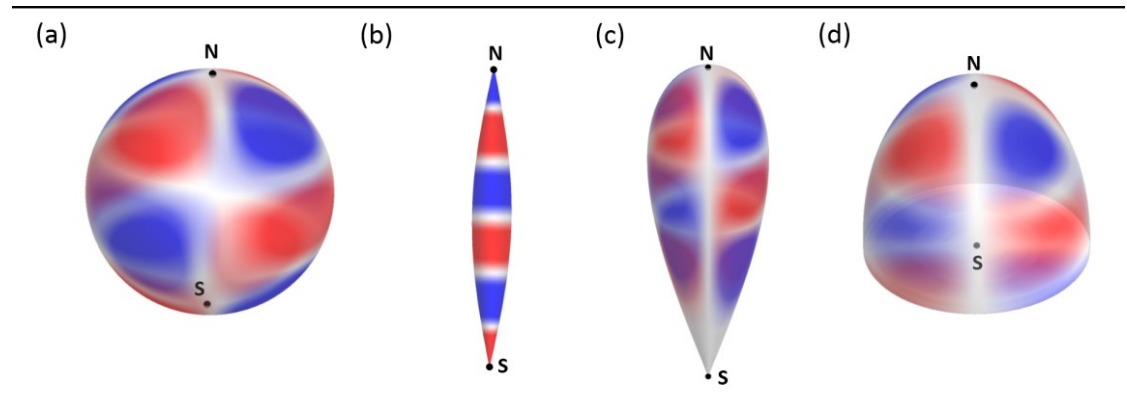

**Figure 2 | Waves at eigen-frequencies in GLs.** Real part of wave functions: (a) $\psi_{52}$ on sphere, (b) $\psi_{50}$ on spindle, (c) $\psi_{52}$ on Tannery's pear, (d) $\psi_{52}$ on truncated Tannery's pear.

The frequency spectra of AIs have a distinctive feature: they are, at least approximately, degenerate and equidistant. This has been shown by solving the Helmholtz equation by numerical methods[6] as well as using the Wentzel–Kramers–Brillouin (WKB) approximation[4, 8]. For GLs, Helmholtz equation in curved 2D surfaces is written as,

$$\tilde{\nabla}_r^2 \psi + \frac{\omega^2}{c^2} \psi = 0, \quad (5)$$

where $\psi$ is wave function, $\omega$ is eigen-frequency and $\tilde{\nabla}$ is Laplacian on curved surface. Because of rotational symmetry of GLs, we can make an ansatz of wave function, namely,

$$\psi(h,\varphi) = R(h)e^{im\varphi}, \quad (6)$$

where $m \in N$ is angular periodic number and $R$ is function depended on $h$. Substituting Eq. (6) to Eq. (5), we get

$$\frac{\partial^2 R}{\partial h^2} + \frac{1}{\rho}\frac{\partial \rho}{\partial h}\frac{\partial R}{\partial h} + \left(\frac{\omega^2}{c^2} - \frac{m^2}{h^2}\right)R = 0. \quad (7)$$

We can solve eigen-problem of Eq. (7) numerically. We find that $\omega$ is (at least approximately) degenerate with $m$. The solutions of Eq. (7) with appropriate boundary conditions can be numbered by a non-negative integer $N$, which represents the number of nodes of the function $R(h)$. Therefore, we can denote wave function and its eigen-frequency with $\psi_{Nm}$ and $\omega_{Nm}$, respectively. These results are summarized in Table 1.

Moreover, $\omega_{Nm}$ can be written as,

$$\omega_{Nm} = a \cdot s_{Nm} + b, \quad (8)$$

where $a$, $b$ are constants and $s_{Nm}$ is an integer for all modes $\psi_{Nm}$. It implies that the spectrum of GLs is (at least approximately) degenerate and equidistant. We plot some typical wave functions on each GLs, as shown in Fig. 2.

Table 1 Description of four AIs and corresponding GLs with spectrum.

| Lens | Refractive index profile | Geodesic lens | Description of geodesic lens | Spectrum |
|---|---|---|---|---|
| Maxwell's fish-eye lens | $n(r) = \dfrac{2}{1+(r/r_0)^2}$ | Sphere | $h(\rho) = \arcsin(\rho)$ | $\dfrac{\omega r_0}{c} = \sqrt{(N+m)(N+m+1)}$ $\approx N + m + 0.5,$ |
| Generalized Maxwell's fish-eye lens | $n(r) = \dfrac{2(r/r_0)^{1/M-1}}{1+(r/r_0)^{2/M}},$ $M = 2,3,4...$ | Spindle | $h(\rho) = \arcsin(\rho),$ $M = 2,3,4...$ | $\dfrac{\omega r_0}{c} = \dfrac{\sqrt{(N+M\cdot m)(N+M\cdot m+1)}}{M}$ $\approx \dfrac{N+0.5}{M} + m,$ |

| extended invisible lens | $(r/r_0)n^{3/2} + (r/r_0)n^{1/2} - 2 = 0$ | Tannery's pear | $h_1(\rho) = -\rho + 2\arcsin(\rho)$,<br>$h_2(\rho) = \rho + 2\arcsin(\rho)$ | $\frac{\omega r_0}{c} \approx \frac{N+0.5}{2} + m$, |
|---|---|---|---|---|
| inverse invisible lens | $(r/r_0)n^{3/2} + (r/r_0)n^{1/2} - 2 = 0; r > r_0$<br>$n = 1; r < r_0$ | Truncated Tannery's pear | $h_1(\rho) = -\rho + 2\arcsin(\rho)$,<br>$h_2(\rho) = 2 + \pi - \rho$ | $\frac{\omega r_0}{c} \approx N + m + 0.5$, |

**Periodic evolution of optical waves and quantum waves in GLs**

Eq. (8) presents a very natural condition. Only if the spectrum has this form, then there exists a period $T$ after which the wave will be the same as at the beginning, up to a possible phase factor. Indeed, expressing a general wave as a superposition of the eigenmodes as $\psi = \sum_{Nm} c_{Nm} \psi_{Nm}$ and employing the mode time evolution $\psi_{Nm}(t) = \psi_{Nm}(0) e^{-i\omega_{Nm} t}$, we find with the help of Eq. (8) that:

$$\psi_{Nm}(T) = e^{-2\pi i b/a} \psi_{Nm}(0), \quad (9)$$

when $T = 2\pi/a$. This way, for an optical wave in GLs, the pulse motion is periodic. This corresponds well to what is known about optical AIs - a pulse emitted from one point will come back to that point after some time, and then it will start spreading again off from that point[3, 8]. Therefore, GLs and their corresponding AIs have similar property of periodic evolution of optical waves. We use a point source to show the periodic evolution of light wave in sphere and truncated Tannery's pear with animations in **Supplementary Information I**. The initial states of animations of sphere and truncated Tannery's pear are shown in Fig. 3(a) and Fig. 3(b), respectively.

**Analogy with quantum mechanics**

There is an interesting similarity of the Helmholtz equation (5) and stationary Schrödinger equation for a free (potential-less) quantum particle on the same geodesic lens. Indeed, the latter equation reads:

$$-\frac{1}{2}\tilde{\nabla}_r \quad -r \quad , (10)$$

where we have used the units where $\hbar$ . If we also set c=1 and relate the optical frequency to the particle energy by $E = \omega^2/2$, the two equations become identical. They therefore have the same solutions, i.e., the same set of eigenmodes. Moreover, it is not hard to show that for all the four geodesic lenses we have discussed, the corresponding quantum frequencies (equal to E in our units) obey the rule (8) as well. Therefore the state of the quantum particle moving on the geodesic lenses will be recovered after a certain time as well, as can be seen in **Supplementary Information II**. However, since the quantum frequencies are proportional to the optical frequencies squared, there will be much more dispersion in the quantum case and quantum waves will change dramatically with time.

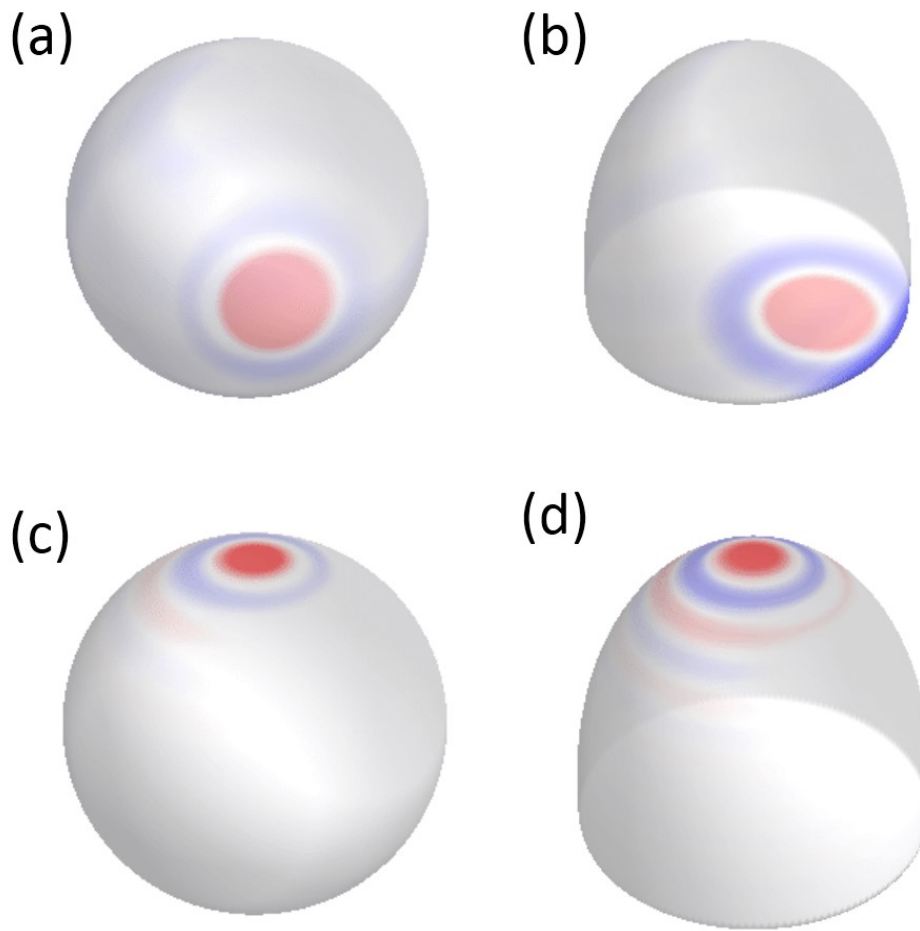

**Figure 3 | The initial states of animations of sphere and truncated Tannery's pear for optical waves in sphere** (a) and truncated Tannery's pear lens (b), for quantum waves in sphere (c) and truncated Tannery's pear lens (d).

**Experimental fabrication of two geodesic lenses**

To verify the closed trajectories in geodesic lenses, we performed two experiments involving light beams propagating on a curved surface in 3D space. One geodesic lens is a spindle corresponding to a generalized Maxwell's fisheye lens with M=5. The other is a sphere corresponding to a Maxwell's fisheye lens. The experiment is achieved based on polymethyl methacrylate (PMMA) waveguide which is fabricated through the self-assembling of polymer solutions on solid substrates [18, 19].

Our geodesic lens is a 2D waveguide with PMMA polymer layer spin-coated on the surface on a metallic needle in spindle shape. The PMMA polymer layer has a uniform thickness of about two micrometers and is doped with $Eu^{3+}$ ions to emit fluorescent light at 615 nm when illuminated with a laser beam with 460 nm wavelength. The optical measurement is based on waveguide excitation and fluorescent imaging techniques shown in figure 4(a) [20]. In the process, the beam is coupled through grating from top of the sample as shown in Fig. 4(a) and propagates in the waveguide. The fluorescent light goes through a color filter and finally is collected by a CCD camera. Figure 4(b) displays the structure of the waveguide in details. We can take it as a bulk material with a refractive index around 1.52 on the surface. We use spin-coating method to deposit the PMMA polymer layer on the surface of metallic needle shape. The details of the sample fabrication can be found in **Supplementary Information III.** Figure 4(c) and 4(d) show the scanning electron microscope image of the metallic needle shape with different scales before spin-coating. Based on these figures, we can accurately obtain the parameters of the metallic needle shape. In Fig. 4(c), the height of the part of spindle which we observed under Scanning Electron Microscope (SEM) is 644 μm, and its diameter ranges from 169 μm to 346 μm. The coupling grating, i.e., the cross structure shown in Fig. 4(c), was fabricated before the spin-coating process.

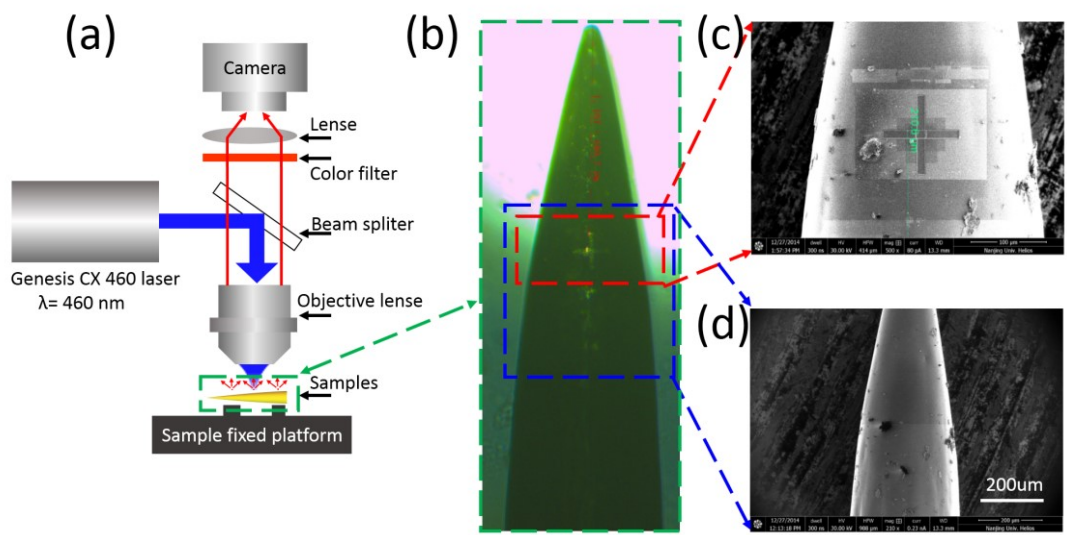

**Figure 4 | Experimental set-up and sample description.** (a) Schematic of the observation and coupling scheme of the light to the geodesic lens. A laser beam is coupled to a 3D curved waveguide from the top, and excites the rare earth ions in the waveguide. The emitted fluorescent light at 615 nm is then collected by a CCD camera. (b) A 3D curved waveguide morphology captured by a CCD camera when illuminated by white light. (c) Scanning electron microscope image of the 3D curved surface around the coupling grating (red dash box) before spin-coating. The cross structure is corresponding to that displayed in (b) and is used to couple laser beam into the waveguide. (d) Scanning electron microscope image of the 3D curved surface with larger scale (blue dash box) before spin-coating. Based on this figure, one can get the accurate parameters of the 3D curved surface.

In Fig. 5(a), we find that a spindle with M=5 in Fig. 1(f) is exact the shape of our metallic needle in

Fig. 4(d). The edge of our sample is almost part of the outline of the spindle (in blue). A point source in spindle has a closed trajectory in red. Fig. 5(b) shows that a laser hitting on the cross grating generates a coupling source on a metallic needle shape, which further results in a light trajectory on a curved surface by using CCD camera. An enlarged drawing nearby the coupling source is shown in Fig. 5(c). The details of experimental measurements can be found in **Supplementary Information IV**. We use the spindle in Fig. 1(a) to fit the measurement of Fig. 5(b). The observed light trajectory matches the calculated one very well, as shown in Fig. 5(d). Based on the results of the shape using SEM and light trajectories using CCD camera, we confirm that we have fabricated a part of spindle with M=5 and observed a part of a closed trajectory.

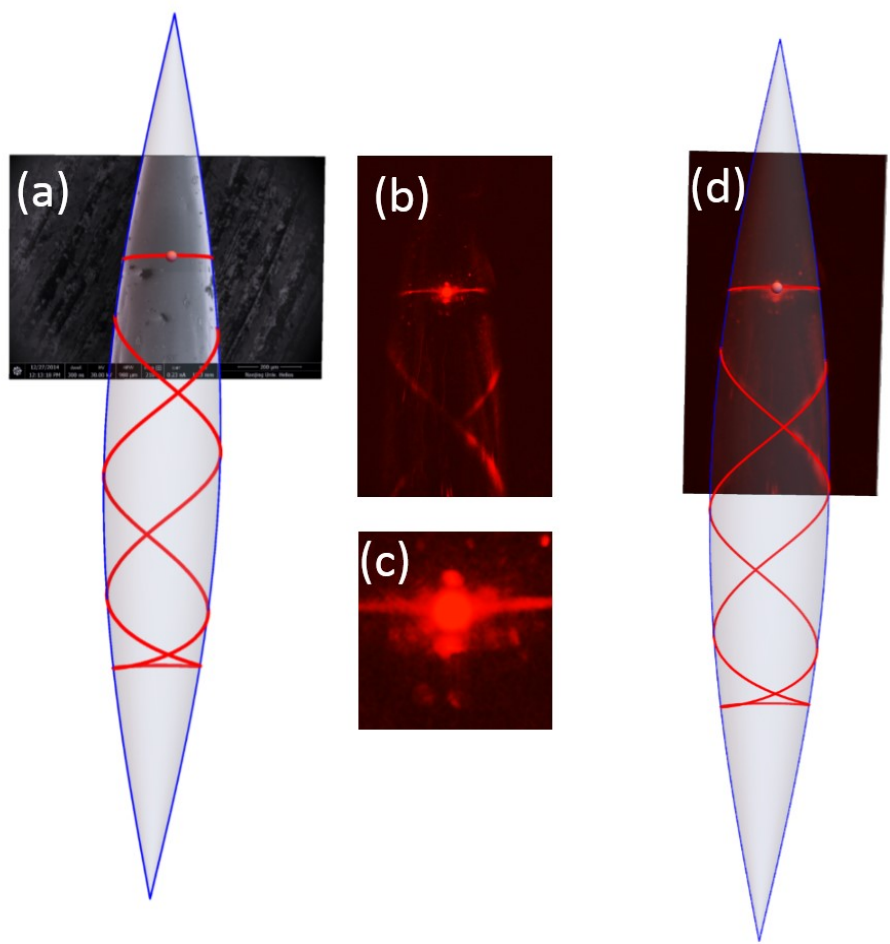

**Figure 5 | Optical measurements and fitting results of light rays in a spindle with M=5.** (a) Fitting the shape of micro-structured metallic needle waveguide with a spindle. (b) Optical measurements on micro-structured metallic needle waveguide. (c) Fitting the light trajectory of micro-structured metallic needle waveguide with a spindle.

Similarly, we also fabricate a sphere as shown in Fig. 6(a) under CCD camera. The sphere is made of iron and its radius is 1 millimeter. By spin-coating method, a uniform PMMA polymer layer is formed surround on the sphere. A handle is attached to the sphere to fix it. A grating was etched nearby the handle to couple laser beam into the waveguide, which results in closed trajectories on sphere. By turning of the background illumination in Fig. 6(a), a clearer picture of closed light trajectory on sphere is shown in Fig. 6(b). We also compare the experimental measurement in Fig. 6(b) with the shape of Fig. 1(e), which has demonstrated that a sphere has been fabricated and a closed light trajectory is observed. It is noted that light rays without closed trajectory on sphere made of BK7 glass had been reported[19].

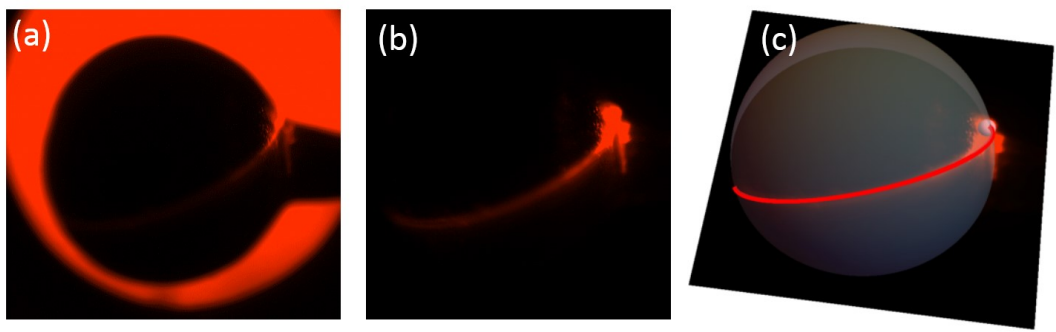

**Figure 6 | Optical measurements and fitting results of light rays on a sphere.** (a) The CCD camera picture of micro-structured sphere waveguide. (b) The light trajectory on micro-structured sphere waveguide. (c) Fitting the light trajectory with a sphere.

To conclude, we have demonstrated that the light trajectories of GLs are closed and the spectrum of compact GLs is (at least approximately) degenerate and equidistant with numerical methods. With this, we demonstrate a periodical evolution of optical waves and quantum waves in GLs. We further fabricate two GLs, namely, a spindle and a sphere in sub-micrometer scale. In these two GLs, we observe curved closed light trajectories with high accurate precision. Our results may offer a new platform to investigate light propagation in curved surfaces[21, 22, 23, 24].

**Acknowledgements**
National Science Foundation of China for Excellent Young Scientists (grant no. 61322504); Foundation for the Author of National Excellent Doctoral Dissertation of China (grant no. 201217); Fundamental Research Funds for the Central Universities (Grant No. 20720170015); National Key Projects for Basic Researches of China (No. 2017YFA0205700 and No. 2017YFA0303700); National Natural Science Foundation of China (No. 11690033, No. 61322504, No. 11621091, No. 61425018, and No. 11374151); L. X. was supported by the China Scholarship Council for half-year study at Masaryk University. T. T. was supported by Grant No. P201/12/G028 of the Czech Science Foundation.

**Author contributions**
L.X., T.T. and H.C. derived the theory, X.W., H.L. , C.S. and S.N.Z. designed and performed the experiments and analysed the data. We co-wrote the paper. L.X., X.W. and T.T. contribute equally to this work.

## Competing interests

The authors declare no competing financial interests.

## Additional information

Correspondence and requests for materials should be addressed to T.T., H.L., H.C.

## References

1. Born M, Wolf E. *Principles of optics: electromagnetic theory of propagation, interference and diffraction of light*. CUP Archive, 2000.

2. Luneburg RK, Herzberger M. *Mathematical theory of optics*. Univ of California Press, 1964.

3. Tyc T, Herzánová L, Šarbort M, Bering K. Absolute instruments and perfect imaging in geometrical optics. *New J. Phys.* **13,** 115004 (2011).

4. Tyc T, Danner AJ. Absolute optical instruments, classical superintegrability, and separability of the Hamilton-Jacobi equation. *Phys. Rev. A* **96,** 053838 (2017).

5. Pendry JB. Negative refraction makes a perfect lens. *Phys. Rev. Lett.* **85,** 3966 (2000).

6. Tyc T, Danner A. Frequency spectra of absolute optical instruments. *New J. Phys.* **14,** 085023 (2012).

7. Tyc T, Chen H, Danner A, Xu Y. Invisible lenses with positive isotropic refractive index. *Phys. Rev. A* **90,** 053829 (2014).

8. Tyc T. Spectra of absolute instruments from the WKB approximation. *New J. Phys.* **15,** 065005 (2013).

9. Zuzaňáková K, Tyc T. Scattering of waves by the invisible lens. *J. Optics* **19,** 015601 (2016).

10. Rinehart R. A solution of the problem of rapid scanning for radar antennae. *J. Appl. Phys.* **19,** 860-862 (1948).

11. Cornbleet S, Rinous P. Generalised formulas for equivalent geodesic and nonuniform refractive lenses. IEE Proceedings H-Microwaves, Optics and Antennas; 1981: IET; 1981. p. 95.

12. Šarbort M, Tyc T. Spherical media and geodesic lenses in geometrical optics. *J. Optics* **14,** 075705 (2012).

13. Leonhardt U. Perfect imaging without negative refraction. *New J. Phys.* **11,** 093040 (2009).

14. Perczel J, Tyc T, Leonhardt U. Invisibility cloaking without superluminal propagation. *New J. Phys.* **13,** 083007 (2011).

15. Leonhardt U, Philbin T. *Geometry and light: the science of invisibility*. Dover Inc. Mineola New York, 2010.

16. Besse AL. *Manifolds all of whose geodesics are closed*, vol. 93. Springer Science & Business Media, 2012.

17. ŠARBORT M. *Non-Euclidean Geometry in Optics, PHD thesis*. Masaryk University 2013.

18. Sheng C., Liu H., Wang Y., Zhu S. N., Genov D. Trapping light by mimicking gravitational lensing. Nature Photonics 7, 902 (2013).

19. Sheng C., Bekenstein R., Liu H., Zhu S.N., Segev M. Wavefront shaping through emulated curved space in waveguide settings. Nature Communications 7: 10747 (2016).

20. Wang X., Chen H. Y., Liu H., Xu L., Sheng C., Zhu S. N. Self-Focusing and the Talbot Effect in Conformal Transformation Optics. Phys. Rev. Lett. 119, 033902 (2017).

21. Schultheiss VH, Batz S, Szameit A, Dreisow F, Nolte S, Tünnermann A*, et al.* Optics in curved space. *Phys. Rev. Lett.* **105,** 143901 (2010).

22. Schultheiss VH, Batz S, Peschel U. Hanbury Brown and Twiss measurements in curved space. *Nat. Photonics* **10,** 106-110 (2016).

23. Bekenstein R, Kabessa Y, Sharabi Y, Tal O, Engheta N, Eisenstein G*, et al.* Control of light by curved space in nanophotonic structures. *Nat. Photonics* **11,** 664 (2017).

24. Bekenstein R, Nemirovsky J, Kaminer I, Segev M. Shape-Preserving Accelerating Electromagnetic Wavepackets in Curved Space. *Physical Review X* **4,** (2014).

# Supplementary Information for:
# Light rays and waves on geodesic lenses

I.  Animations of the periodic evolution of light waves in sphere (A1) and truncated Tannery's pear (A2) .

II. Animations of the periodic evolution of quantum waves in sphere (A3) and truncated Tannery's pear (A4).

### III. Sample fabrication

A spindle waveguide with uniform thickness was fabricated for our experiment, as depicted in Fig. 4b in the main text. The sample is a polymethyl methacrylate (PMMA) resist polymer layer on the surface on a metallic spindle. The fabrication process is based on metallic wire fusion technique and spin-coating technique. Firstly, a straight silver wire with a diameter of 400 micrometers was put on a hydrogen flame (Fig. S1a). Secondly, the ends of the straight silver wire were pulled at a specific speed and broken into two parts (Fig. S1b and S1c). Thirdly, the broken end (Fig. 4c and 4d in the main text) of the silver wire was coated with PMMA layer, and the rare earth ions $Eu^{3+}$ was added to the PMMA resist to facilitate fluorescence imaging that would reveal the propagation dynamics of light beams. Finally, a three dimensional metallic waveguide, namely, a polymer layer on the surface of the broken end of the silver wire was obtained (Fig. 4d in the main text). In the experiment, the shape of the metallic spindle could be changed by controlling the speed of pulling the ends of the straight metallic wire, the materials of the metallic wire and the distance of the silver wire and the hydrogen flame. The polymer layer thickness and uniformity can be varied by changing the solubility of the PMMA solution, the evaporation rate of the oven and the pulling rate, immersion rate, and cycling time of the spin-coating. By changing these parameters, a spindle waveguide of the desired shape was fabricated. The thickness of the waveguide is large enough such that we effectively behaves as a bulk material with a refractive index around 1.52. A sketch of the waveguide is shown in Figure S2.

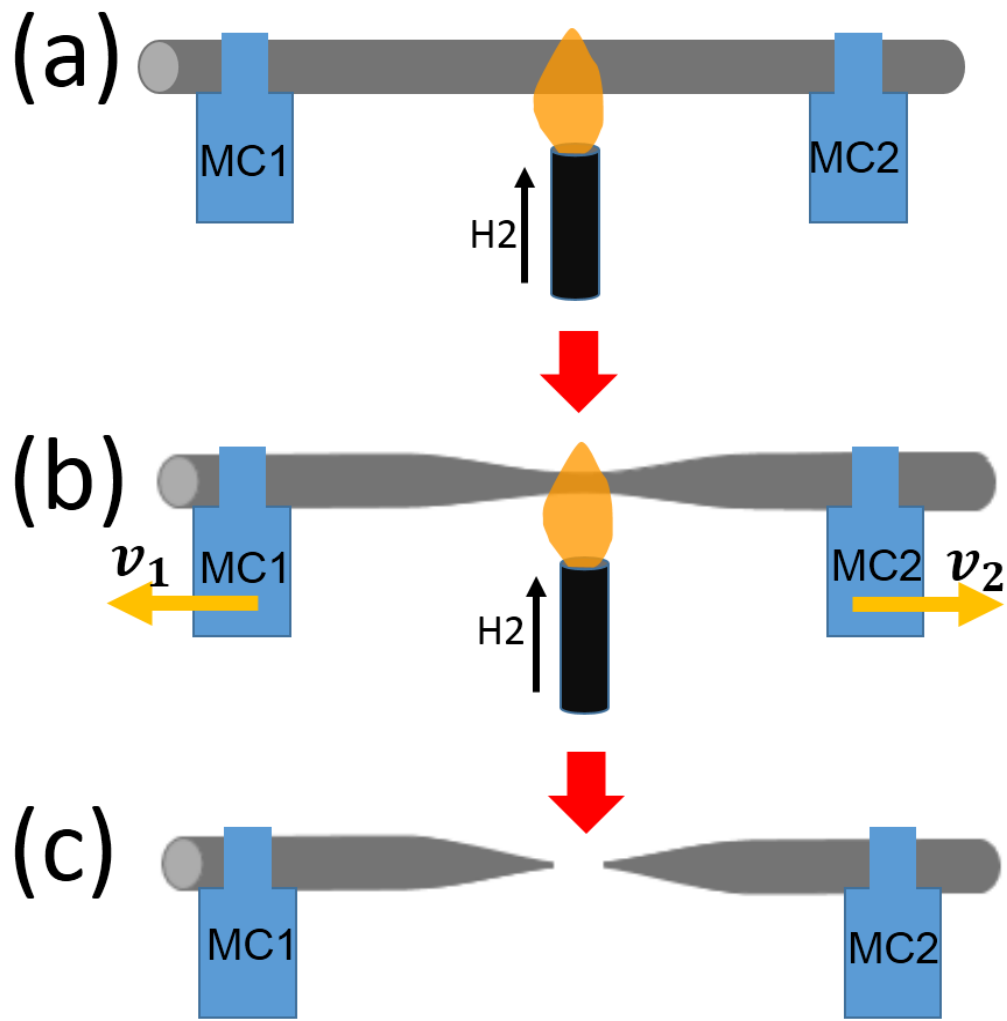

Figure S1. Sample fabrication process. (a) The position of straight silver wire, movement console(MC) and hydrogen flame. A straight silver wire was fixed on MC1 and MC2, and then was put on a hydrogen flame at a proper position. (b) Metallic wire fusion process. The ends of the straight silver wire were pulled at speeds ($v_1$ and $v_2$), and the silver wire gradually became tapered. (c) Two obtained metallic cones. The silver wire was broken into two metallic cones at some point.

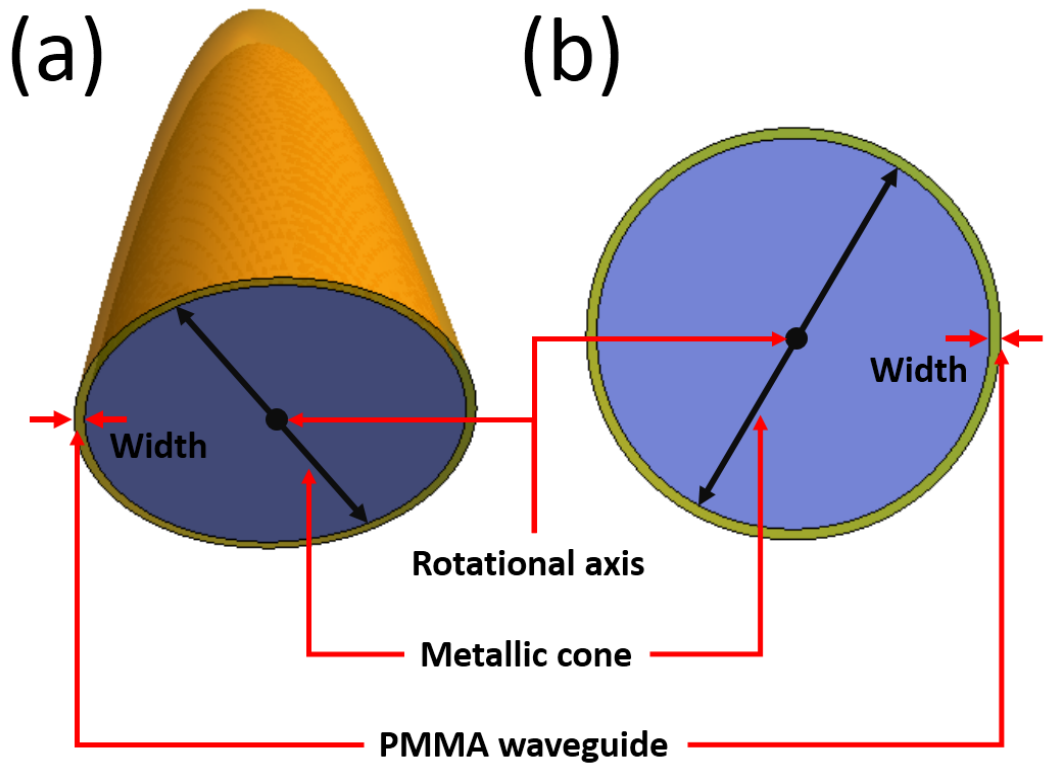

Figure S2. (a) Sketch of a metallic waveguide. (b) Cross section of the metallic waveguide.

IV. **Experimental measurements**

As mentioned in the above Sample fabrication section, a metallic waveguide was generated along the metallic curved surface by means of spin-coating technology. We then use the experimental setup (Fig. 4a in the main text) to measure the light propagation in the metallic waveguide. A laser beam with at a wavelength of 460nm was coupled to the waveguide by a grating with a period of 310 nm, which was fabricated on the metallic curved surface with focused ion beam (FEI Strata FIB 201, 30 keV, 11 pA) before the spin coating process. To better illustrate the coupling process, we make a sketch of the grating (Figure S3 (a)). It (in the yellow boxes) is corresponding to the red dash box in Figure S3 (b), which is an SEM of the curved surface. After spin-coating process, a thick enough PMMA polymer layer was deposited on the surface. The blue laser beam is perpendicularly incident on the grating, and converted to the PMMA waveguide. Figure S3 (c) shows the grating coupling process and the optical measurement of the metallic waveguide.

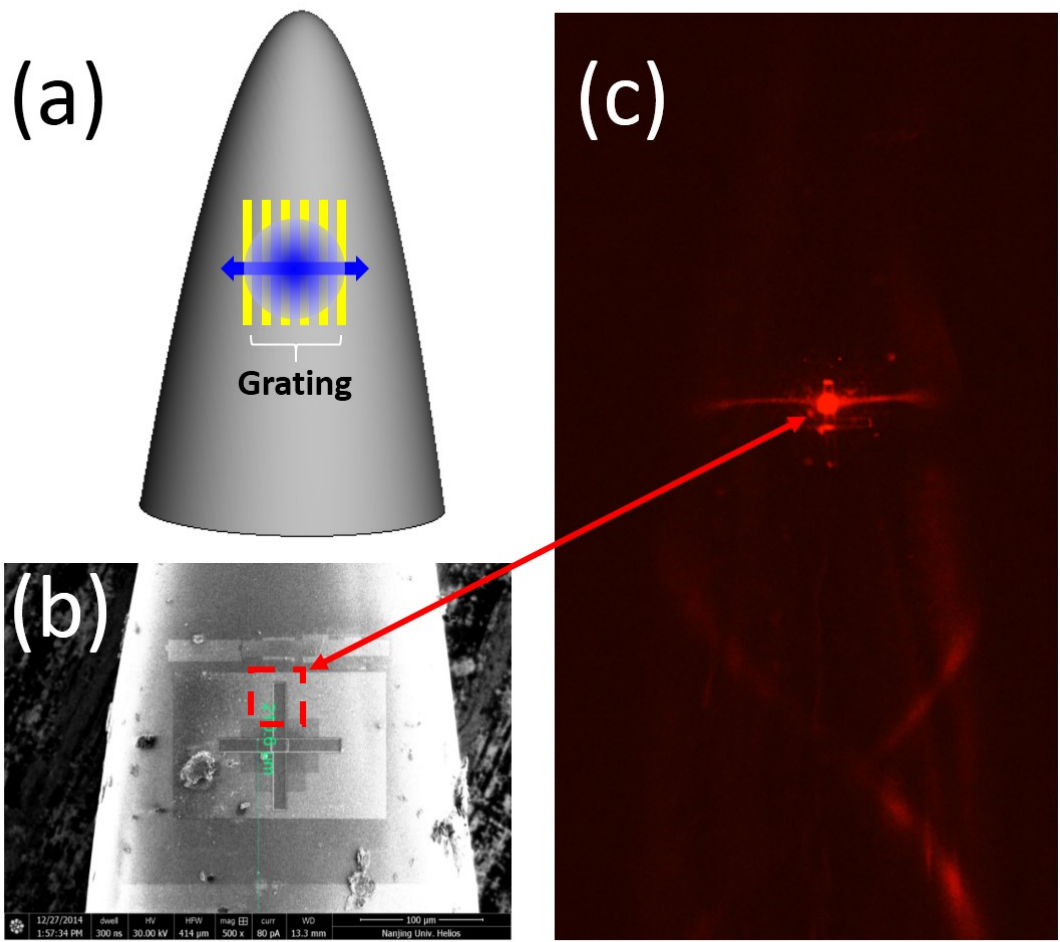

Figure S3. (a) Sketch of the coupling grating (yellow boxes). (b) SEM of the metallic curved surface and the coupling grating (in red dash box) before spin-coating process. (c) The grating coupling process and optical measurement of the sample. The yellow boxes in (a) show the coupling grating, the graded blue spot represents the exciting beam, and the blue arrows show the direction of the laser beam propagating in the waveguide. The coupling gratings in (a) is corresponding to the red dash box in (b), which is fabricated before spin-coating process. The red dash box region in (b) is corresponding to the grating in (c), which is indicated by the red arrows.